# Larger photovoltaic effect and hysteretic photocarrier dynamics in Pb[(Mg$_{1/3}$Nb$_{2/3}$)$_{0.70}$Ti$_{0.30}$]O$_3$ crystal

A S Makhort, G Schmerber and B Kundys[1]

Université de Strasbourg, CNRS, Institut de Physique et Chimie des Matériaux de Strasbourg, UMR 7504, 23 rue du Loess, F-67000 Strasbourg, France

Keywords: ferroelectrics, photovoltaics, photopolarization

**Abstract**
Following the recent discovery of a bulk photovoltaic effect in the Pb[(Mg$_{1/3}$Nb$_{2/3}$)$_{0.68}$Ti$_{0.32}$]O$_3$ crystal, we report here more than one order of magnitude improvement of photovoltaicity as well as its poling dependence in the related composition of lead magnesium niobate-lead titanate noted Pb[(Mg$_{1/3}$Nb$_{2/3}$)$_{0.7}$Ti$_{0.30}$]O$_3$. Photocurrent measurements versus light intensity reveal a remarkable hysteresis in photocarrier dynamics clearly demonstrating charge generation, trapping and release processes.

## 1. Introduction

Ultimately approaching fundamental limit [1, 2] of semiconductor photovoltaic (PV) technology stimulates the development of an alternative to p-n junction-based solar energy conversion. One of the promising routes can be found in electrically polar materials where non-zero intrinsic electric field can replace p-n region of semiconducting photovoltaic cells with an ability to generate above bandgap photovoltages [3, 4]. Indeed, electrically polar photovoltaic materials have gain renewed attention in photovoltaics [5–11] and related multi-functionalities [12–21]. Although photovoltaic effect in non-centrosymmetric crystals have long been known [22], renewed attention occurred after the discovery of photovoltaic effects in the multiferroic BiFeO$_3$ [23, 24] with recent progress in photovoltaic efficiency of Bi$_2$FeCrO$_6$ films, reporting a record value of 8.1% [25]. Because the bulk photovoltaic effect (BPVE) can be modified by extrinsic contributions (i.e. possible surface/interface effects in films [26] or grain size dependence in ceramics [27]) investigations on single crystals offer unique fundamental insight into photoelectric property. Among the still scarce PV compounds the initially non-photovoltaic ferroelectric (FE) Pb[(Mg$_{1/3}$Nb$_{2/3}$)$_{0.64}$Ti$_{0.36}$]O$_3$ (PMN-PT36%) crystals were reported to exhibit the PV effect after doping with WO$_3$ [28]. Although this compound belongs to the well-known family of piezoelectric crystals with multifunctional phase diagram, other members of this composition were not tested until recently [29]. Since a much larger photovoltaic effect was found even in the undoped PMN-PT32% crystal with a composition closer to the morphotropic phase boundary, a careful study of other compounds in the phase boundary region occurring between PT = 30 and PT = 35% [30] becomes very promising. Here we report the existence of a photovoltaic effect in the PMN-PT30% compound which exceeds by more than 1 order of magnitude the effect reported for the PMN-PT32% counterpart. We further compare photovoltaic and ferroelectric performances of the two compositions and report unprecedented light-induced charge dynamics responsible for this effect.

## 2. Experimental details

The crystals had (001) orientation supplied by Crystal-Gmbh (Germany) in square shape with edges along [010] and [100] directions (figure 1(a) (inset)). Electrodes were formed with silver paste covering the edges in the planes parallel to *zy*. The hysteresis loop of polarization versus electric field was taken at room temperature by using a quasi-static FE loop tracer similar to that described in [31], reducing FE fatigue by ultralow frequency (0.01 Hz) measurements. The sample was illuminated with a 365 nm (3.4 eV) UV-LED with 30 nm spectral linewidth with an intensity of 153 mW cm$^{-2}$ in order to investigate the change in the FE polarization response. The current was monitored by a Keithley electrometer (Model 6517B) at the time constant of 0.36 s. The temperature of the sample measured by a thermal camera (Therm-app) increases by ⩽1.9 K under light illumination and such temperature change made no noticeable difference to the FE loop.

---

[1] Author to whom any correspondence should be addressed (kundysATipcms.unistra.fr)

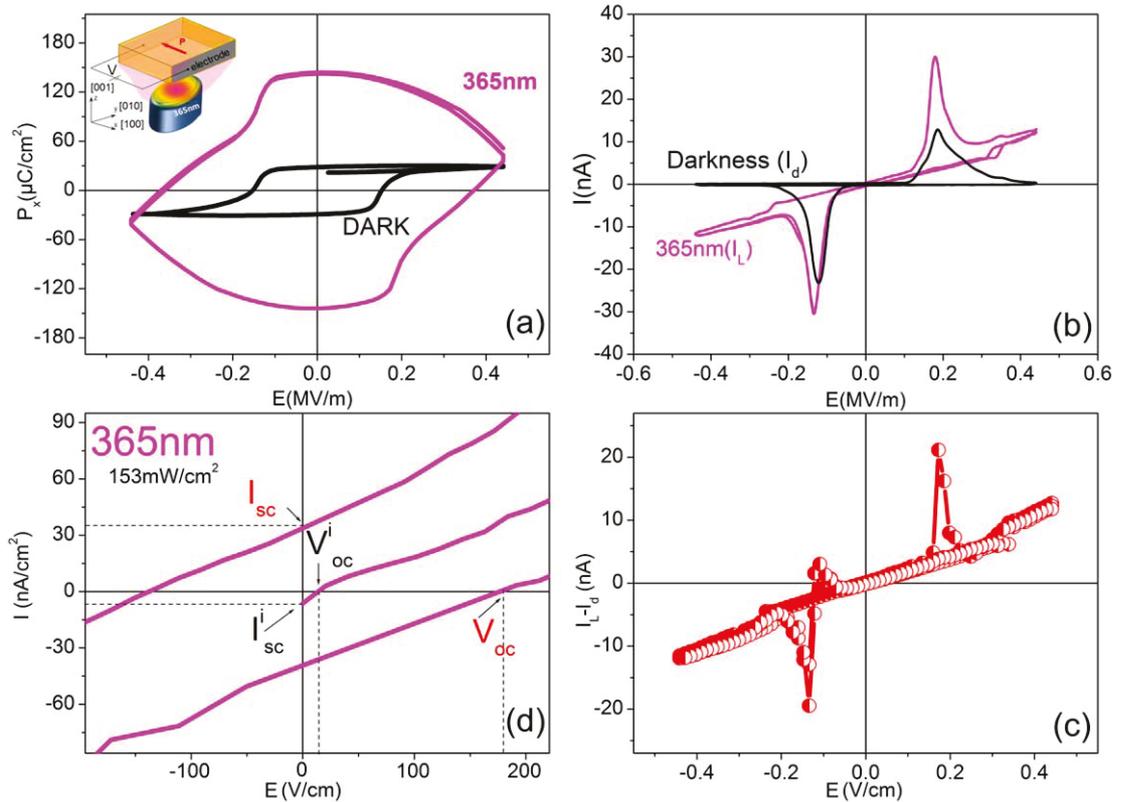

**Figure 1.** Ferroelectric polarization (a) and current loops (b) in darkness and under light (365 nm, 153 mW cm$^{-2}$). Inset to figure 1(a) shows schematics of the experiment. Figure (c) represents the difference between FE currents under light and in darkness. Figure (d) shows a zoomed evolution the ferroelectric current under light in the remanent polarization state with ±1kVcm$^{-1}$ electric field amplitude.

## 3. Results and discussions

The FE loop measured in darkness along [100] direction reveals a classical hysteresis behavior resulting in the two polarization states of about ±29 $\mu$C cm$^{-2}$, in agreement with the literature data (figure 1(a)). However when illumination of 365 nm light is applied the apparent polarization increases by more than 4 times.

As a consequence of free charge generation by light, the sample becomes leakier FE with apparent increase in the both FE polarization and FE coercive force (figure 1(b)). The observed light-induced change in the ferroelectric loop largely exceeds in magnitude all previous observations [29, 32–34]. The corresponding volt-ampere characteristics further illustrate the photoinduced change in electric properties (figure 1(b)). The main three effects arise under light illumination: (i) the significant increase of current related to dipole reorientation (ferroelectric peak); (ii) the general increase in the sample conductivity; and (iii) a noticeable shift in abscissa of the FE loop as a result of light generated charges contribution to the total intrinsic electric field. The difference between FE currents in darkness and under illumination is presented in figure 1(c). As it can be seen the maximum of the light induced effect is achieved at the ferroelectric peak that is strongly poling history dependent in ferroelectrics. The change in the basic parameters of the photovoltaic effect is better illustrated in the zoomed region (figure 1(d)). The initial values of short circuit photocurrent $I_{sc}^i$ and open circuit photovoltage $V_{oc}^i$, increase largely after poling with ±1 kV cm$^{-1}$. In particular, there is also a noticeable down shift of the loop along $y$-axis, so the absolute value of $I_{sc}$ after pooling with +1 kV cm$^{-1}$ is smaller than the value of $I_{sc}$ obtained after pooling with −1 kV cm$^{-1}$. Same effect is also seen for open circuit photovoltage $V_{oc}$. Because $I_{sc}$ and $V_{oc}$ are used to evaluate photovoltaic efficiency, the electric tuning becomes possible. These extraordinary properties were more clearly observed by us in the photovoltaic Bi$_2$FeCrO$_6$ films [35] and can be expected to be a general and technologically important electrically switchable feature for photovoltaic ferroelectric compounds.

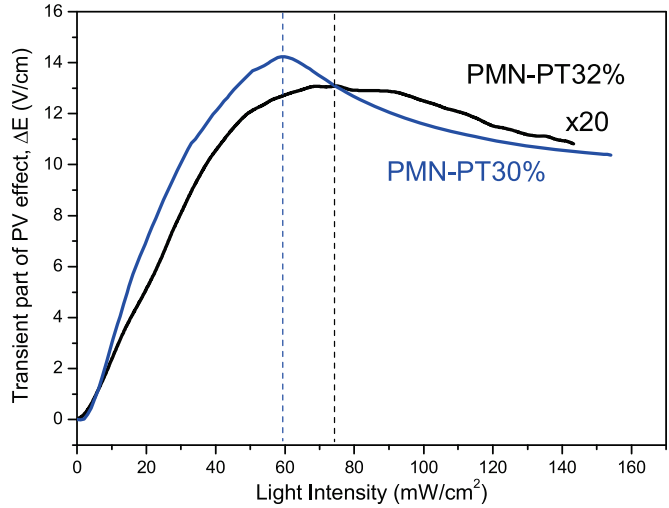

**Figure 2.** Comparison of photovoltaic effect for the Pb[(Mg$_{1/3}$Nb$_{2/3}$)$_{0.68}$Ti$_{0.32}$]O$_3$ (32%) and Pb[(Mg$_{1/3}$Nb$_{2/3}$)$_{0.70}$Ti$_{0.30}$]O$_3$ (30%) crystals at room temperature for samples in a remanent FE state.

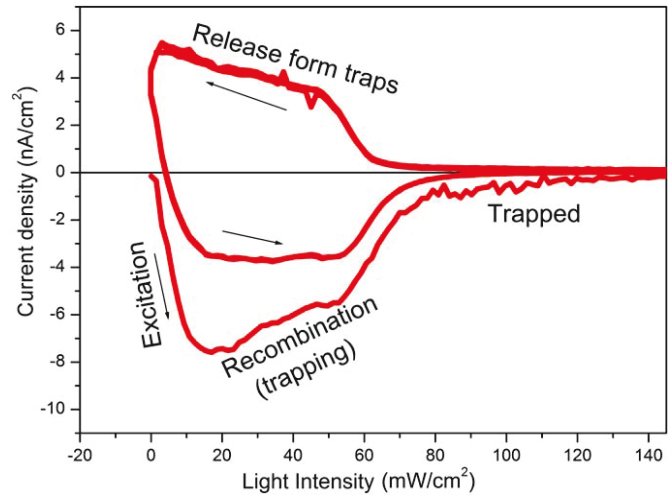

**Figure 3.** Photocurrent as a function of light intensity for Pb[(Mg$_{1/3}$Nb$_{2/3}$)$_{0.70}$Ti$_{0.30}$]O$_3$(30%) crystals at room temperature.

At ferroelectric remanence the voltage change versus light intensity shows a ~20-times larger effect than in the PMN-PT32% crystal (figure 2) in agreement with much larger effect of the light on the FE loop (figure 1(a) and [29]). The nonlinear behavior as a function of light intensity with a characteristic peak is observed for both compounds. The form of curves can be explained by the occurrence of two competing mechanisms: the light induced charge generation dominant at low light intensities and the charge recombination processes at higher intensities (figure 2). These opposing processes give rise to the peak as a function of light intensity at the value where numbers of generated and recombined carriers are expected to become comparable. Notably, the maximum photovoltaic effect is reached faster for the PMN-PT30% (~59 mW cm$^{-2}$) than for PMN-PT32% (74 mW cm$^{-2}$). In order to get insight into the origin of the observed behavior we have measured the related electric current as a function light intensity (so-called Lux-Ampere-like characteristic (figure 3)). Prior to measurements the sample was set to the remanent polarization state by sweeping the electric field from $-0.4$ MV m$^{-1}$ to $+0.4$ MV m$^{-1}$ and then to zero, to ensure a monodomain configuration. The light intensity was then increased and the current was monitored by a Keithley electrometer (Model 6517B) with a related time constant of 0.36 s. The observed behavior can be tentatively explained as follows. The charges, initially generated by light, move in the previously defined polarization direction, and therefore create a current (linear part,

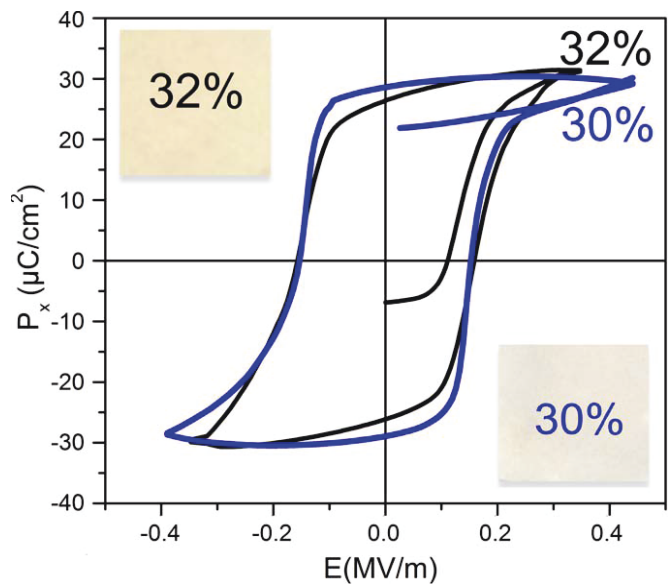

**Figure 4.** Comparison of ferroelectric loops in darkness between Pb[(Mg$_{1/3}$Nb$_{2/3}$)$_{0.68}$Ti$_{0.32}$]O$_3$(32%) and Pb[(Mg$_{1/3}$Nb$_{2/3}$)$_{0.70}$Ti$_{0.30}$]O$_3$(30%) crystals. Insets shows pictures of the both crystals on white background indicating larger absorption in the case of Pb[(Mg$_{1/3}$Nb$_{2/3}$)$_{0.68}$Ti$_{0.32}$]O$_3$(32%) (yellow color).

figure 3). This light-induced charge generation process eventually reaches its saturation and the photo-carriers start to recombine reducing the related current to zero (trapped state, figure 3).

As light intensity decreases, this process is reversed, and the majority of carriers become free again, but now move in the opposite direction, in agreement with the sign change of the photocurrent (see 'release from traps' part in figure 3). Subsequent illumination reveals a smaller magnitude of the photocurrent, because a part of photo-excited carriers had recombined irreversibly, diminishing surface charges, and therefore decreasing polarization [29]. The role of polarization magnitude in the PV effect can be verified by comparing ferroelectric loops of both compounds in darkness (figure 4). The PMN-PT30% compound with a larger photovoltaic effect indeed shows a slightly larger remanent polarization.

However, the difference of 0.2 $\mu$C cm$^{-2}$ is rather too small to be responsible for the large difference in the photovoltaic properties reported in figure 2. The same argument can be pointed out for pyroelectric coefficients that do not change much for Pb[(Mg$_{1/3}$Nb$_{2/3}$)$_x$Ti$_{1-x}$]O$_3$ composition for $0.3 < x < 0.32$ [36]. Another explanation can be based on the possible difference in the optical absorption coefficients between the two compounds. However, the optical absorption is known to decrease with decreasing $x$ in the Pb[(Mg$_{1/3}$Nb$_{2/3}$)$_x$Ti$_{1-x}$]O$_3$ composition [37] in agreement with the fact that the Pb[(Mg$_{1/3}$Nb$_{2/3}$)$_{0.70}$Ti$_{0.30}$]O$_3$ (PT30%) crystals are more transparent at ambient conditions (figure 4 (insets)). On the other hand, the composition of Pb[(Mg$_{1/3}$Nb$_{2/3}$)$_x$Ti$_{1-x}$]O$_3$ crystals is optimized to obtain large piezoelectric properties, reaching a maximum for $x = 0.3$, exactly at the lower morphotropic phase boundary [30]. Thus, thanks to the enhanced piezoelectric properties in the PMN-PT30% compound, the light-generated charges can contribute more efficiently to the electric field-assisted transformation between the thermodynamically equivalent phases at the morphotropic phase boundary. The light then induces changes in polarization that are connected to stress. Consequently, the larger is the piezoelectric coefficient, the larger light-induced effect is expected on the lattice deformation [38], which, in turn can modify the bandgap [39–41] of the material leading to the increased photovoltaic effect. Although this mechanism is possible, its contribution is unlikely dominant here because the piezoelectric coefficients do not differ by the order of magnitude in both compounds [30]. The intrinsic mechanism of light induced charge generation may come into play deserving a separate study including symmetry dependent [29] and defect dependent [42] arguments.

## 4. Conclusions

In conclusion, an enhancement in the photovoltaic effect of more than one order of magnitude has been found in the Pb[(Mg$_{1/3}$Nb$_{2/3}$)$_{0.70}$Ti$_{0.30}$]O$_3$ compound, with the composition at the lower border of the morphotropic phase diagram. The much larger photovoltaic effect at the lower border of the MPB and unprecedented Lux-Ampere-like characteristic demonstrating hysteretic photo carrier dymamics for the first time should be regarded as key basic findings. This study should rapidly prompt a screen of other compounds of the same family as well as similar compositions [43–45] in which photovoltaic effects can occur thanks to acentricity [46] with the aim to better understand and optimize their photovoltaic properties.